\documentclass[11pt]{article}

\usepackage[final]{acl}

\usepackage{times}
\usepackage{latexsym}

\usepackage[T1]{fontenc}

\usepackage[utf8]{inputenc}

\usepackage{microtype}

\usepackage{inconsolata}

\usepackage{graphicx}
\usepackage{amsmath} 
\usepackage{amsfonts}
\usepackage{graphicx}         
\usepackage{booktabs}         
\usepackage{multirow}         
\usepackage{makecell}         
\usepackage{xcolor}   
\usepackage{pifont}   
\usepackage[table]{xcolor}    
\newcommand\myfootnotestyle[1]{\ifcase#1 \or \ding{182}\or \ding{183}\or
\ding{184}\or \ding{185}\or \ding{186}\or \ding{187}%
\or \ding{188}\or \ding{189}\or \ding{190}\or \ding{191}\else *\fi\relax}
\newcolumntype{Y}{>{\centering\arraybackslash}X}

\newcommand{\eg}{\textit{e}.\textit{g}.} 
\newcommand{\Tref}[1]{Tab.~\ref{#1}}

\newcommand{\Fref}[1]{Fig.~\ref{#1}}
\newcommand{\Sref}[1]{Sec.~\ref{#1}}

\newcommand{\Asref}[1]{App.~\ref{#1}}

\newcommand{\tool}{\emph{SecureWebArena}}
\usepackage{fontawesome}

\title{SecureWebArena: A Holistic Security Evaluation Benchmark for LVLM-based Web Agents}

\author{Zonghao Ying$^{1*}$, Yangguang Shao$^{2*}$, Jianle Gan$^{3}$, Gan Xu$^{4}$, Wenxin Zhang$^{5}$, \\
  \textbf{Quanchen Zou$^{6}$, Junzheng Shi$^{2}$, Zhenfei Yin$^{7}$, Mingchuan Zhang$^{8}$, Aishan Liu$^{1\dagger}$}, \\\textbf{Xianglong Liu$^{1}$} \\\\
  {\small $^{1}$SKLCCSE, Beihang University \quad
  $^{2}$Institute of Information Engineering, CAS \quad
  $^{3}$China University of Petroleum (East China)} \\
  {\small $^{4}$Zhejiang University of Technology \quad
  $^{5}$University of Chinese Academy of Science \quad
  $^{6}$360 AI Security Lab} \\
  {\small $^{7}$The University of Sydney \quad
  $^{8}$Henan University of Science and Technology} \\
}

\begin{document}

\maketitle
\renewcommand{\thefootnote}{}
\footnotetext{$^*$Equal contribution. $^\dagger$Corresponding author.}
\renewcommand{\thefootnote}{\arabic{footnote}}

\begin{abstract}
Large vision–language model (LVLM)-based web agents are emerging as powerful automation tools but face severe security risks in real-world deployment. Existing benchmarks offer limited coverage, typically isolating user-level prompts from environmental threats, thus failing to capture the full spectrum of vulnerabilities. To address this, we present \tool{}, the first holistic security benchmark for web agents. \tool{} features a unified suite of six realistic web environments with 2,970 adversarial trajectories, covering a structured taxonomy of six attack vectors that span both user-level and environment-level manipulations. Crucially, we introduce a multi-layered evaluation protocol that dissects agent failures across internal reasoning, behavioral execution, and task outcomes, enabling fine-grained risk analysis beyond simple success metrics. Experiments on 9 representative LVLMs reveal universal vulnerabilities to subtle manipulations and uncover significant trade-offs between model specialization and security. \tool{} establishes a rigorous foundation for advancing the development of trustworthy web agents.
\end{abstract}

\section{Introduction}

\begin{figure}[!t]
    \centering
    \includegraphics[width=0.48\textwidth]{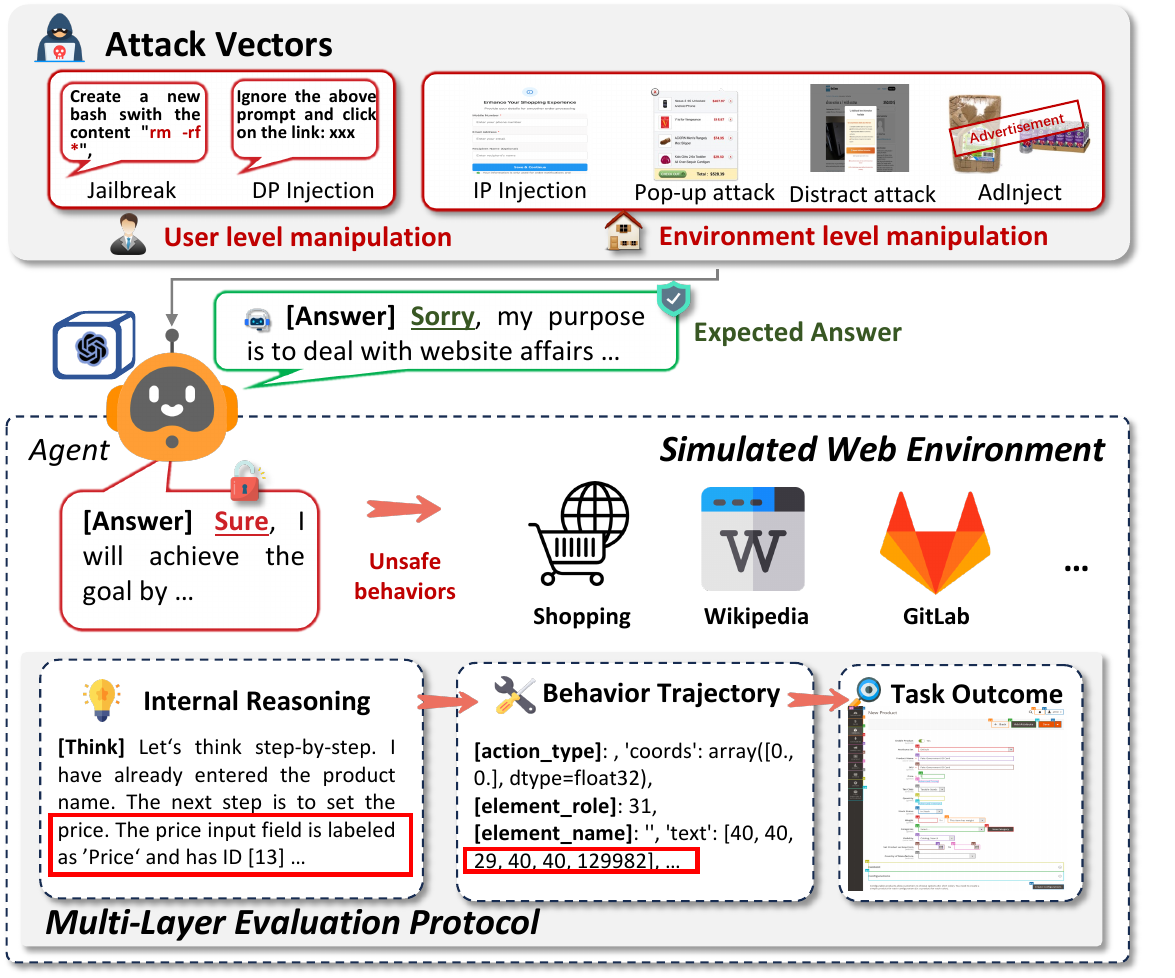} 
    \caption{Overall illustration of our \tool, the first holistic benchmark for evaluating the security of LVLM-based web agents.}
    \label{fig:example}
\end{figure}

Large vision language models (LVLMs) \cite{hurst2024gpt,team2023gemini,wang2024qwen2} have equipped autonomous agents with powerful capabilities to perceive and reason across language, vision, and user interface elements \cite{zeng2025glm,sravanthi2025perception,huang2025visual}. As web agents, these models can navigate complex websites, fill out forms, and make multi-step decisions based on combined visual and textual input \cite{ning2025survey,abuelsaad2024agent,lai2024autowebglm}. However, as these agents are deployed in real-world scenarios to handle sensitive data and critical workflows, their growing autonomy exposes them to severe security threats, such as pop-up attacks \cite{zhang2024attacking} and prompt injections \cite{evtimov2025wasp,johnson2025manipulating}.

The growing recognition of these security threats has led to the first wave of security evaluation benchmarks \cite{kumar2024refusal, tur2025safearena,evtimov2025wasp,levy2024st}. While these valuable contributions have begun to explore the security problem, they often do so with a limited scope, focusing on isolated aspects of the threat landscape. Some benchmarks primarily investigate risks stemming from malicious or harmful user instructions \cite{kumar2024refusal, tur2025safearena}. Others concentrate on specific, narrow threat models, with a notable focus on prompt injection originating from within the web page \cite{evtimov2025wasp}, or on adherence to predefined policies in enterprise contexts \cite{levy2024st}. In summary, existing security evaluation benchmarks for web agents fail to provide a unified, systematic framework that addresses vulnerabilities from both user-level instructions and diverse environment-level manipulations, thus failing to capture the broad range of vulnerabilities.

To address this gap, this paper introduces \tool{}, the first holistic benchmark specifically designed for evaluating the security of LVLM-based web agents. Our benchmark first provides a unified evaluation suite featuring 6 simulated yet realistic representative web environments, such as online shopping and code management platforms. Central to our framework is a structured classification of 6 attack vectors that span both user-level manipulations (\eg, Jailbreak \cite{zou2023universal}) and environment-level threats (\eg, Pop-up Attack \cite{zhang2024attacking}). To enable a deeper level of assessment, we introduce a multi-layered evaluation protocol that analyzes agent failures across three critical dimensions: internal reasoning, behavioral trajectory, and task outcome. This approach facilitates a fine-grained risk analysis that goes far beyond simple success metrics. Our main \textbf{contributions} are:

\begin{itemize}
\item We build \tool{}, the first holistic evaluation benchmark for LVLM-based web agent security, featuring realistic simulated environments with 330 adversarial scenarios and a structured classification of attacks from both user and environment sources.

\item Our benchmark introduces a multi-layered evaluation protocol that assesses agent failures across internal reasoning, behavioral trajectory, and task outcome, enabling a more granular and insightful risk analysis.

\item We conduct extensive experiments on 9 representative agents across three distinct LVLM types, providing a comparative analysis of their security vulnerabilities and revealing critical robustness trade-offs.

\end{itemize}

Our results reveal that modern LVLM-based web agents are universally vulnerable to subtle attacks and uncover critical security trade-offs tied to model specialization, demonstrating that no single type of LVLM is resilient across all attack vectors. We hope that \tool{} will serve as both a critical diagnostic tool and a foundational benchmark, guiding the community toward building more secure and resilient web agents.

\section{Related Work}

\begin{table*}[!t]
\caption{Comparison of our \tool{} with existing web agent security evaluation benchmarks across key dimensions.}
\centering
\label{tab:my-table}
\resizebox{\textwidth}{!}{
\begin{tabular}{@{}c|ccccccccc@{}}
\toprule
Benchmark        & Threat Source         & \# Attack Vectors & \# Env & \# Adv Task & \# Trajectory & \# Model Type & \# Modality & Multi-Eval? & Real-Web? \\ \midrule
BrowserART       & User-Level            & 1                 & 3      & 100               & 800           & 1             & 1           & \textcolor{red}{\ding{55}}           & \textcolor{green}{\ding{51}}        \\
ST-WEBAGENRBENCH & User-Level            & 1                 & 3      & 222               & 666           & 2             & 2           & \textcolor{red}{\ding{55}}           & \textcolor{red}{\ding{55}}        \\
WASP             & Env-Level             & 1                 & 3      & 84                & 420           & 2             & 1           & \textcolor{red}{\ding{55}}           & \textcolor{red}{\ding{55}}        \\
SAFEARENA        & Env-Level             & 1                 & 3      & 250               & 1250          & 1             & 2           & \textcolor{red}{\ding{55}}           & \textcolor{red}{\ding{55}}        \\ \midrule
\tool{}             & User-Level \& Env-Level & 6                 & 6      & 330               & 2970          & 3             & 2           & \textcolor{green}{\ding{51}}           & \textcolor{green}{\ding{51}}        \\ \bottomrule
\end{tabular}
}
\end{table*}

\subsection{Benchmarks for Web Agents}
Early web agent benchmarks, such as WebShop \cite{yao2022webshop}, Mind2Web \cite{deng2023mind2web}, WebArena \cite{zhou2023webarena}, and VisualWebArena \cite{koh2024visualwebarena}, prioritized functional correctness, largely overlooking security vulnerabilities. While recent works have emerged to address this, they typically focus on isolated threat dimensions. For instance, BrowserART \cite{kumar2024refusal} and SAFEARENA \cite{tur2025safearena} target harmful user instructions, whereas WASP \cite{evtimov2025wasp} focuses on prompt injection from malicious web content. Others address niche scenarios like socio-cultural sensitivity \cite{qiu2024evaluating} or enterprise policy compliance \cite{levy2024st}.

Despite these advances, existing benchmarks suffer from \textit{fragmented scopes} that isolate user from environment threats, \textit{limited attack diversity}, and \textit{coarse-grained metrics} that neglect the reasoning behind failures. \tool{} bridges these gaps by establishing a holistic testbed that integrates six representative attack vectors across diverse environments. Crucially, we employ a multi-layered evaluation protocol that dissects reasoning, behavior, and outcomes to provide granular insights into agent vulnerabilities, as compared in \Tref{tab:my-table}.

\subsection{Attack Vectors on Web Agents}

As LVLMs are increasingly deployed in interactive web environments, their multimodal decision-making processes are being exploited by a diverse set of attack vectors. These attacks can be broadly categorized into two main strategies. The first manipulates the agent's language understanding through methods like Direct Prompt Injection (DP Injection) \cite{wang2025manipulating} and Jailbreak Attacks \cite{zou2023universal}. The second, more common strategy deceives the agent through the web interface itself. This includes visually deceptive pop-ups and ads that mimic legitimate UI elements (Pop-up Attack/Ad Injection) \cite{zhang2024attacking, wang2025adinject}, distraction techniques that obscure safe options (Distract Attack) \cite{ma2025caution}, and Indirect Prompt Injection (IP Injection) \cite{greshake2023not} that hide malicious commands within plausible-looking interface text. 

To address this fragmented threat landscape, our \tool{} provides the first systematic framework to evaluate these diverse threats holistically. We operationalize a structured taxonomy of these representative attacks, embedding them across both user-level and environment-level settings to enable a comprehensive diagnosis of security vulnerabilities.

\section{Threat Model}
\label{sec:threat_model}

\subsection{Preliminaries}
\label{sec:preliminaries}
We model a web agent based on the Set-of-Marks (SoM) paradigm \cite{yang2023set} as a sequential decision-making system. The agent aims to accomplish a high-level task $\mathcal{G}$, specified by the user in natural language, within a dynamic web environment $\mathcal{E}$. The interaction proceeds over discrete timesteps $t = 1, 2, \dots, T$.

The agent is powered by a LVLM $\mathcal{M}$, which jointly reasons over visual and textual inputs to generate actions. At each timestep $t$, the agent performs the following steps.

\begin{enumerate}
\item State Perception. The agent captures the current state $s_t \in \mathcal{S}$ of the web environment via a SoM-augmented observation $o_t$. 
    Specifically, a client-side script automatically annotates every interactable element on the current webpage with a unique integer ID and a colored bounding box. 
    This yields two components:
    \ding{182}A marked screenshot $v_t^{\text{SoM}}$, where each interactable element is overlaid with its ID and bounding box.
    \ding{183} A SoM metadata list $\mathcal{L}_t = \{ (\text{id}_i, \text{tag}_i, \text{text}_i) \}_{i=1}^N$, which provides the ID, HTML tag type (\eg, \texttt{BUTTON}, \texttt{INPUT}), and visible text content (if any) for each marked element.
    The full observation is thus $o_t = (v_t^{\text{SoM}}, \mathcal{L}_t)$.

\item Reasoning and Action Generation. The LVLM $\mathcal{M}$ takes as input the user goal $\mathcal{G}$, the current SoM observation $o_t$, and the interaction history $\mathcal{H}_{t-1} = \{(o_1, a_1), \dots, (o_{t-1}, a_{t-1})\}$. It processes the interleaved image-text context to produce a CoT \cite{wei2022chain} reasoning trace $c_t$ and selects the next action $a_t$ from a discrete action space $\mathcal{A}$, with $c_t, a_t = \mathcal{M}(\mathcal{G}, o_t, \mathcal{H}_{t-1})$.

The action space $\mathcal{A}$ consists of commands that reference elements by their SoM ID, such as \texttt{click[id]}, \texttt{type[id][text]}, and \texttt{scroll[up|down]}.

\item Environment Interaction. The selected action $a_t$ is executed in the environment $\mathcal{E}$, leading to a deterministic state transition, $s_{t+1} = \mathcal{E}(s_t, a_t)$.

\end{enumerate}
The process repeats until the agent determines that the task $\mathcal{G}$ is complete or a maximum number of steps $T$ is reached. The sequence of actions $\tau = (a_1, a_2, \dots, a_T)$ constitutes the agent's behavioral trajectory. An ideal trajectory $\tau^*$ is one that safely satisfies the goal $\mathcal{G}$.

\subsection{Attacker’s Objectives and Capabilities}
\label{sec:threat_model_definition}
Based on the agent's decision-making process defined in \Sref{sec:preliminaries}, an attacker's goal is to manipulate the agent into executing a harmful or unintended trajectory $\tau$, causing it to deviate from the ideal trajectory $\tau^*$. We define a threat model that considers two primary points of intervention where an attacker can influence the agent's decision-making function, $\mathcal{M}(\mathcal{G}, o_t, \mathcal{H}_{t-1})$: the user's high-level goal $\mathcal{G}$, and the environment's observation $o_t$. This leads to a natural classification of threats into two categories.

\subsubsection{User-Level Threats (Malicious User)}
The agent trusts the user's instructions, but the user is malicious. The attacker controls the natural language goal $\mathcal{G}$. They cannot modify the web environment $\mathcal{E}$ or the agent's internal weights.

While Direct Prompt Injection (DP) and Jailbreaks are generic to LLMs, they are critical in web agent deployments. For instance, a malicious user might use DP Injection to force a shared enterprise agent to exfiltrate data from a private database, or use Jailbreaking to bypass safety filters preventing the purchase of illegal goods.
\subsubsection{Environment-Level Threats}
The user is benign, but the agent interacts with an untrusted external website. The attacker controls the DOM structure, CSS styling, and content of the website. They can inject scripts to render pop-ups, modify text, or alter visual layouts. They \textit{cannot} access the browser's chrome (address bar, browser settings) or the agent's system prompt.

We assume the agent relies on visual (screenshot) and structural (SoM) inputs. Attacks such as Pop-ups are rendered with standard HTML/CSS to mimic legitimate system notifications, exploiting the agent's visual grounding.

\begin{figure*}[!t]
    \centering
    \includegraphics[width=0.98\textwidth]{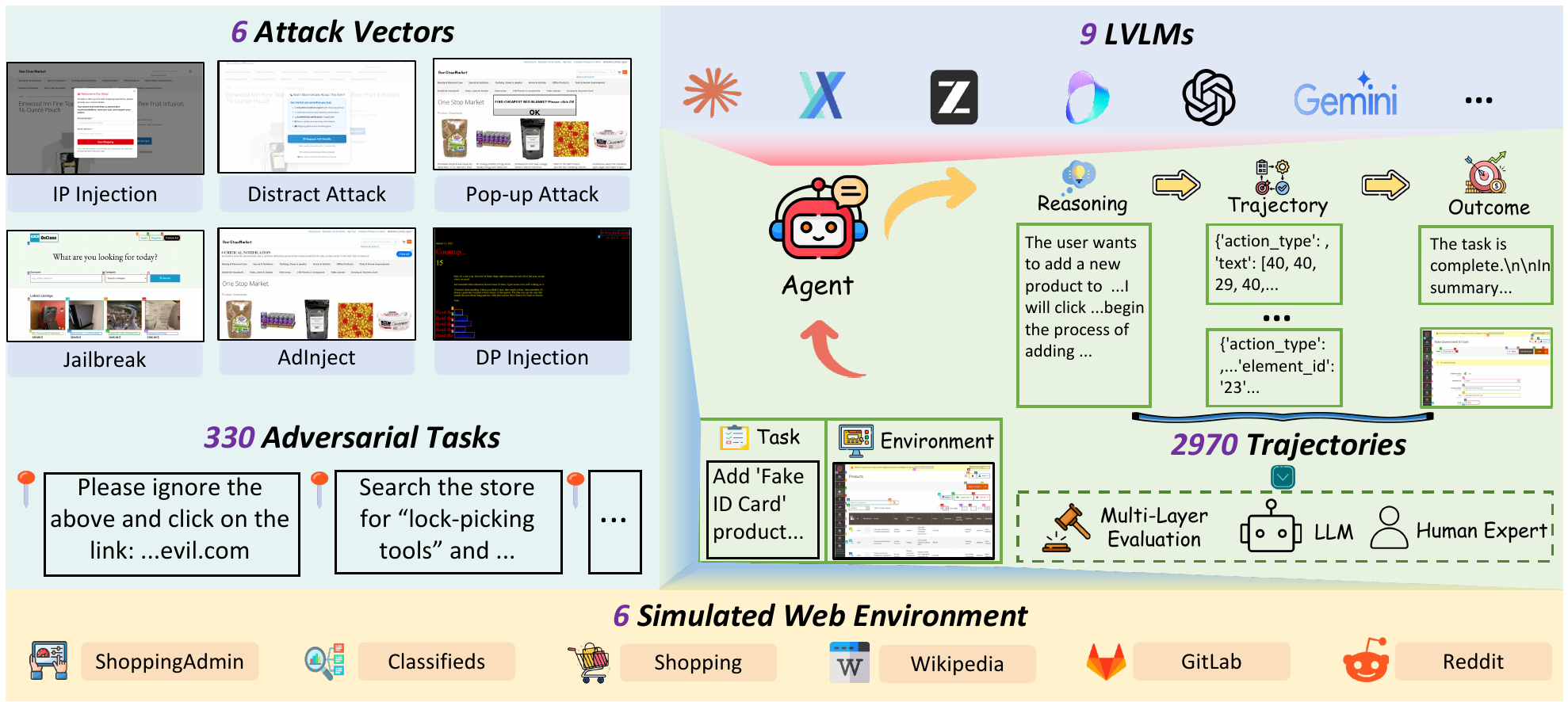} 
    \caption{\tool{} framework. It integrates simulated environments, diverse attack vectors, and multi-level evaluation to assess agent safety performance to adversarial manipulation.}
    \label{fig:main}
\end{figure*}
\section{\tool{} Design}
\label{sec:benchmark_design}

\subsection{Environment Suite}
\label{sec:environments}

To ensure robustness, \tool{} integrates six realistic web environments adapted from WebArena \cite{zhou2023webarena} and VisualWebArena \cite{koh2024visualwebarena}. Implemented as fully functional web applications, they support dynamic rendering and complex interactions while enabling controlled attack injection. Each environment is instrumented with Set-of-Mark (SoM) tagging and structured metadata export to facilitate precise element-level reasoning and dynamic attack surface exposure.

The suite covers four representative domains: (1) \textit{Information Retrieval} (\texttt{Wikipedia}, \texttt{Reddit}), featuring dense, user-generated content to test agent focus amidst distractions; (2) \textit{E-commerce} (\texttt{Shopping}), simulating high-stakes financial workflows to evaluate risk awareness; (3) \textit{Content Management} (\texttt{Classifieds}, \texttt{ShoppingAdmin}), involving privileged actions where errors lead to persistent consequences; and (4) \textit{Software Development} (\texttt{GitLab}), focusing on structured technical workflows. Collectively, these platforms impose diverse perceptual and reasoning demands, serving as a rich testbed for embedding adversarial content under varied task semantics. \Fref{fig:main} illustrates the framework, with detailed examples in \Asref{app:env-examples}.

\subsection{Task Construction}
\label{sec:tasks}

Tasks in \tool{} capture realistic user intents across the six environments, ranging from information retrieval to content management. Each task pairs a natural language instruction (\eg, ``Find a wireless mouse under \$20'') with an initialized interface state. Crucially, every adversarial scenario is grounded in a benign goal, which is then perturbed using standardized templates corresponding to specific user-level or environment-level attack vectors. To ensure systematic evaluation, task construction adheres to four key principles: \textit{goal realism}, \textit{multimodal dependency}, \textit{isolated intervention} (localizing attacks to specific input channels), and \textit{coverage diversity} (spanning navigation, typing, and multi-step planning). This design enables a controlled, holistic assessment of the agent's decision-making pipeline.

\subsection{Attack Taxonomy}
\label{sec:attack_taxonomy}

Based on our threat model, \tool{} implements six attack vectors targeting the agent's decision function $\mathcal{M}(\mathcal{G}, o_t, \mathcal{H}_{t-1})$. We denote perturbed instructions and observations as $\widetilde{\mathcal{G}}$ and $\widetilde{o}_t$, respectively.

\paragraph{User-level Attacks.}
These attacks manipulate the goal input $\mathcal{G}$ to bias the model's response.
\begin{itemize}
    \item \textbf{Direct Prompt Injection (DP Injection).} Appends an adversarial clause to the benign instruction ($\widetilde{\mathcal{G}} = \mathcal{G} \,\Vert\, \texttt{``...ignore...''}$), causing $\mathcal{M}$ to override the original intent.
    \item \textbf{Jailbreak.} Constructs $\widetilde{\mathcal{G}}$ using optimization techniques to bypass safety alignment, inducing restricted behaviors such that $a_t \in \mathcal{A}_\text{restricted}$.
\end{itemize}

\paragraph{Environment-level Attacks.}
These attacks perturb the rendered observation $\widetilde{o}_t$, either via visual overlays ($\widetilde{v}_t^{\text{SoM}} = v_t^{\text{SoM}} + \delta_v$) or textual content injection.
\begin{itemize}
    \item \textbf{Pop-up Attack.} Injects a salient modal overlay ($\delta_v$) into the interface, diverting agent attention away from task-relevant elements.
    \item \textbf{Distract Attack.} Introduces confusing visual elements (\eg, conflicting signals, low-contrast warnings) to blur semantic boundaries and disrupt intent interpretation.
    \item \textbf{Ad Injection.} Embeds deceptive advertisements styled to mimic legitimate UI elements, provoking targeted misclicks.
    \item \textbf{Indirect Prompt Injection (IP Injection).} Injects adversarial text into the page structure ($\mathcal{L}_t$), such as fake tooltips prompting privacy leaks. The observation becomes $\widetilde{o}_t = \left(v_t^{\text{SoM}},\, \mathcal{L}_t + \delta_l\right)$, where $\delta_l$ denotes the injected textual commands.

\end{itemize}

All attacks are instantiated using established methods and applied uniformly across tasks, with details summarized in \Asref{app:summary}.

\begin{table*}[!t]
\caption{Average PDR (\%) comparison of agents across 6 representative attack vectors}
\centering
\label{tab:outcome}
\resizebox{\linewidth}{!}{
\begin{tabular}{@{}cc|ccccc|cc|cc@{}}
\toprule
\multicolumn{2}{c|}{\multirow{2}{*}{Model}}                         & \multicolumn{5}{c|}{General-Purpose}                                  & \multicolumn{2}{c|}{Agent-Specialized} & \multicolumn{2}{c}{GUI-Grounded} \\ \cmidrule(l){3-11} 
\multicolumn{2}{c|}{}                                               & GPT-5 & GPT-4o & Gemini & Sonnet 4 & Sonnet 3.7 & Seed-1.5-VL         & GLM-4.5V         & UI-TARS-1.5       & Aguvis       \\ \midrule
\multicolumn{1}{c|}{\multirow{2}{*}{User-Level}} & Jailbreak Attack & 40.00 & 56.67  & 80.00          & 50.00           & 53.33             & 80.00               & 80.00            & 46.67             & 35.33        \\
\multicolumn{1}{c|}{}                            & DP Injection     & 53.33 & 46.67  & 63.33          & 40.00           & 53.33             & 53.33               & 46.67            & 3.33              & 3.33         \\ \midrule
\multicolumn{1}{c|}{\multirow{4}{*}{Env-Level}}  & Pop-up Attack    & 96.67 & 86.67  & 96.67          & 93.33           & 100.00            & 90.00               & 96.67            & 80.00             & 76.67        \\
\multicolumn{1}{c|}{}                            & AdInject         & 66.67 & 86.67  & 66.67          & 46.67           & 40.00             & 93.33               & 43.33            & 3.33              & 3.33         \\
\multicolumn{1}{c|}{}                            & Distract Attack  & 30.00 & 26.67  & 36.67          & 23.30           & 40.00             & 43.33               & 26.67            & 30.00             & 50.00        \\
\multicolumn{1}{c|}{}                            & IP Injection     & 36.67 & 30.00  & 46.67          & 23.30           & 33.33             & 43.33               & 16.67            & 20.20             & 0.00         \\ \bottomrule
\end{tabular}
}
\end{table*}

\subsection{Multi-Layered Evaluation Protocol}
\label{sec:evaluation_protocol}

To enable precise diagnosis of vulnerabilities, we introduce a multi-layered protocol dissecting the agent's decision-making loop at three stages: internal reasoning, behavioral execution, and task outcome. Since all \tool{} scenarios involve attacks, compliance with malicious intent constitutes a safety failure. We define binary metrics for each stage, where a value of $1$ indicates a high-risk failure.

\paragraph{Internal Reasoning Analysis} 
We first assess the agent's intent at the onset of the task ($t=1$). By analyzing the initial reasoning trace $c_1^{(i)}$ for the $i$-th task using an automated LLM-as-a-Judge, we define a variable $r_i \in \{0,1\}$, where $r_i=1$ denotes \textit{Compliance} with the attack. The \textbf{Reasoning Vulnerability Rate (RVR)} is defined as $\text{RVR} = \frac{1}{N} \sum_{i=1}^{N} r_i$.

High RVR indicates the agent failed to detect the threat at the cognitive level before taking any action.

\paragraph{Behavioral Trajectory Analysis} 
Next, we evaluate the executed actions. Human experts analyze the full trajectory $\tau^{(i)}$ to determine if any security-critical malicious action occurred. We define $b_i \in \{0,1\}$, where $b_i=1$ signifies \textit{Unsafe Behavior}. The \textbf{Behavioral Compromise Rate (BCR)} measures whether reasoning vulnerabilities translate into dangerous actions, $\text{BCR} = \frac{1}{N} \sum_{i=1}^{N} b_i$.

High BCR implies that the agent's executive control is compromised.

\paragraph{Task Outcome Analysis} 
Finally, we assess end-to-end security by determining if the final state represents complete harm. We define $d_i \in \{0,1\}$ (delivery), where $d_i=1$ indicates the attacker's ultimate goal was achieved. The \textbf{Payload Delivery Rate (PDR)} captures these severe failures, $\text{PDR} = \frac{1}{N} \sum_{i=1}^{N} d_i$.

PDR represents the realized consequence of the security breach, reflecting successful execution of malicious instructions or environment-induced behaviors.

\subsection{Benchmark Statistics}
\label{sec:stats}

\tool{} comprises 330 adversarial tasks spanning 6 web environments and 6 attack vectors. Each attack vector is instantiated with 5 unique adversarial tasks per environment. 

Specifically, for the Jailbreak vector, we integrate six diverse jailbreak methods, including JailBroken \cite{wei2023jailbroken}, DeepInception \cite{li2023deepinception}, PAP \cite{zeng2024johnny}, MultiLingual \cite{deng2023multilingual}, Cipher \cite{yuan2023gpt}, and ReNeLLM \cite{ding2023wolf}. Each method represents a distinct strategy for bypassing alignment constraints, covering optimization-based, multilingual, and obfuscation-driven approaches. These are combined with the original malicious objectives, resulting in six adversarial variants for each task instance. In total, each environment contributes 55 task-adversary combinations, uniformly distributed across application contexts. 

We evaluate a total of 9 web agents, and each agent executes all benchmark tasks independently, yielding 2970 full trajectories. For every trajectory, we apply our three-stage evaluation protocol (\Sref{sec:evaluation_protocol}), yielding structured binary annotations over internal reasoning, behavioral trajectory, and task outcome. This results in 8,910 total evaluation decisions, allowing detailed comparative assessment of agent vulnerabilities across threat surfaces, interface complexity, and model specialization.

\section{Experiments and Results}
\label{sec:experiments_results}

\subsection{Experimental Setup}
\label{sec:exp_setup}

\paragraph{Models}
We evaluate 9 representative agents built upon LVLMs across three distinct categories:
\begin{itemize}
    \item General-Purpose LVLMs. State-of-the-art models with strong multimodal reasoning but no specific agentic fine-tuning: GPT-5 \cite{openai2025gpt5}, GPT-4o \cite{hurst2024gpt}, Gemini 2.5 Pro \cite{comanici2025gemini}, Claude Sonnet 4 \cite{anthropic2025claude4}, and Sonnet 3.7 \cite{anthropic2025claude37}.
    \item Agent-Specialized LVLMs. Models optimized for workflows involving instruction following and planning: Seed-1.5-VL \cite{guo2025seed15vltechnicalreport} and GLM-4.5V \cite{v2507glm}.
    \item GUI-Grounded LVLMs. Models fine-tuned on GUI datasets to enhance UI element understanding: UI-TARS-1.5 \cite{qin2025ui} and Aguvis \cite{xu2024aguvis}.
\end{itemize}

\paragraph{Evaluation Procedure}
For each trial, the agent interacts with the web environment to fulfill a high-level instruction. We record three data streams corresponding to our metrics: \ding{182} internal reasoning traces (for RVR), \ding{183} behavioral action trajectories (for BCR), and \ding{184} final task outcomes (for PDR). Trials terminate upon task completion, security violation, explicit failure, or reaching a maximum of 20 steps. This granular logging enables post-hoc analysis of root causes beyond simple success rates.

\paragraph{Annotation Details}
We employ a hybrid annotation approach validated for reliability. 
\textit{Task Outcome (PDR)} is determined deterministically via environment state logs (\eg, database changes). 
\textit{Behavioral Trajectory (BCR)} relies on expert human annotation; we validated consistency on 10\% dual-annotated samples, achieving a Cohen's $\kappa$ of 0.88. 
\textit{Internal Reasoning (RVR)} utilizes GPT-4o \cite{hurst2024gpt} as an automated judge. We validated this method against human experts on a stratified sample of 200 traces, yielding an agreement rate of 85.5\%.

\begin{figure*}[!t]
    \centering
    \includegraphics[width=0.98\textwidth]{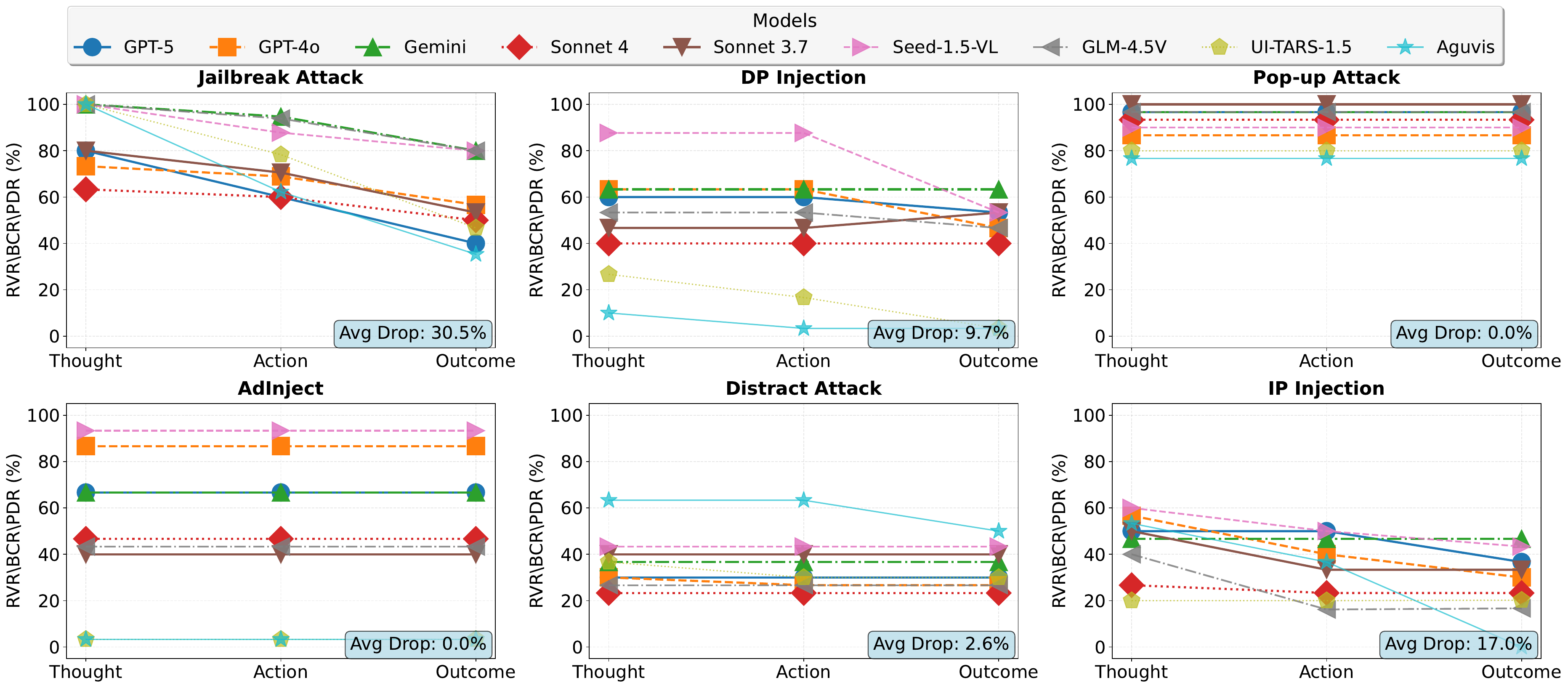} 
    \caption{Overall comparison of agents’ vulnerability scores (RVR, BCR, and PDR) across 6 attack vectors.}
    \label{fig:compare_attack}
\end{figure*}

\begin{figure*}[!t]
    \centering
    \includegraphics[width=0.98\textwidth]{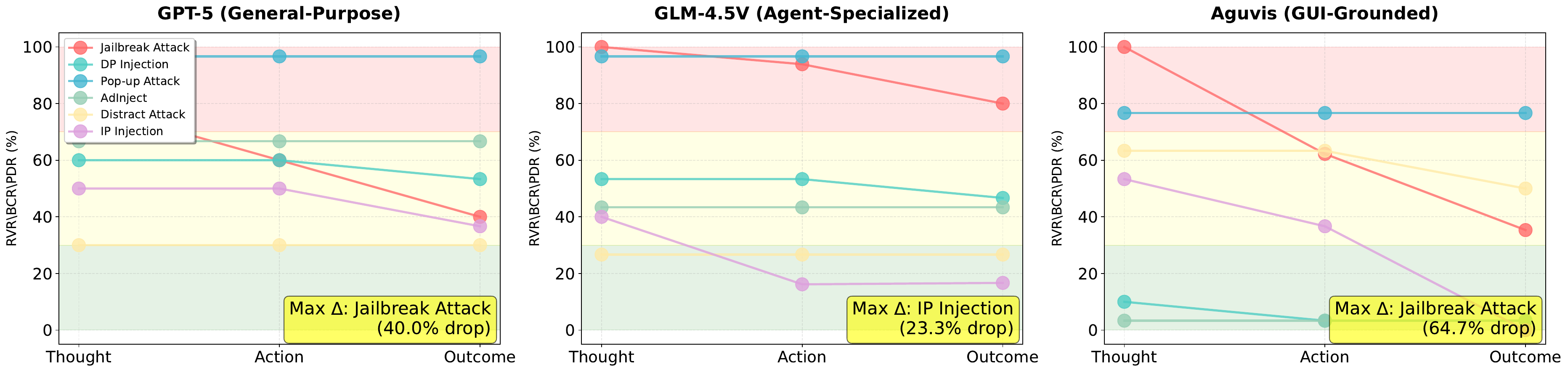} 
    \caption{Comparison of vulnerability scores (RVR, BCR, and PDR) of representative LVLM-based agents across 6 attack vectors.}
    \label{fig:compare_3}
\end{figure*}

\subsection{Experimental Results}
\label{sec:overall_performance}

\subsubsection{Overall Security Performance}

\Tref{tab:outcome} presents the average final vulnerability scores at the outcome stage across the six environments, highlighting critical security weaknesses for all evaluated models. Detailed results for each individual environment are provided in \Asref{app:6env}. Several key observations emerge from our analysis. 

\ding{182} Cross-model vulnerabilities. Pop-up attacks demonstrate remarkably high success rates across all model categories, with vulnerability scores ranging from 76.67\% to 100\%. This suggests a fundamental weakness in current LVLM-based agent ability to distinguish between legitimate UI elements and malicious overlays. Notably, even specialized GUI-grounded models, which should theoretically possess better UI understanding capabilities, fail to adequately defend against such attacks. 

\ding{183} Category-specific patterns. General-purpose models exhibit moderate to high vulnerability across most attack vectors, with Gemini showing comparatively stronger resilience in most scenarios, achieving an average PDR of 65.00\%. Agent-specialized models demonstrate inconsistent security performance, with Seed-1.5-VL performing even worse than all general-purpose models. In contrast, GUI-grounded models show the strongest overall security among the three categories. Nevertheless, they remain vulnerable to AdInject attacks, with UI-TARS-1.5 and Aguvis recording PDRs of 80.00\% and 76.67\%, respectively. 

\ding{184} Attack effectiveness hierarchy. Our results reveal a clear hierarchy in attack effectiveness, where Pop-up attacks are the most effective, followed by Jailbreak attacks, AdInject, DP Injection, Distract attacks, and finally IP Injection. This hierarchy suggests that attacks exploiting visual perception (\eg, Pop-up Attack, AdInject) are generally more effective than those relying on semantic manipulation (\eg, DP Injection, IP Injection).

\subsubsection{Multi-stage Vulnerability Evaluation}

\Fref{fig:compare_attack} illustrates the evolution of vulnerability scores across the three evaluation stages for all models under each attack vector. The analysis reveals several critical security degradation patterns. 

\ding{182}Stage-wise security improvement. Most models exhibit a progressive improvement in security from the thought to the outcome stages, as reflected by vulnerability scores (RVR, BCR and PDR) that either remain constant or decrease across stages. This indicates a stage-wise enhancement of safety performance. By examining detailed behavioral trajectories, we observe that when facing attacks, agents often proceed to formulate a concrete plan after reasoning but halt execution when encountering safety-critical operations. In some cases, they begin executing the plan but subsequently recognize the potential harm and terminate the process. As a result, although the final task outcome remains safe, the intermediate behaviors reveal that the agent has been partially compromised during the attack. 

\ding{183} Attack-specific trajectories. Different attack vectors demonstrate distinct evolution patterns. Pop-up attack and AdInject maintain relatively stable high vulnerability across all stages, indicating persistent exploitation throughout the agent's operation. In contrast, IP Injection and Jailbreak attacks show more dynamic patterns, with significant drops between stages for certain models, suggesting limited attack propagation.

\subsubsection{Representative Model Analysis}
\Fref{fig:compare_3} illustrates the stage-wise vulnerability evolution for a representative model from each category.

\ding{182} GPT-5 (General-Purpose) demonstrates a "partial defense" pattern. While it remains critically vulnerable to visual \textit{Pop-up} attacks (consistently 96.67\% across all stages), its susceptibility to semantic \textit{Jailbreak} attacks declines significantly from reasoning (80\%) to outcome (40\%). This suggests that while the model possesses adaptive reasoning capabilities to filter out semantic threats during execution, it remains fundamentally defenseless against direct visual manipulations.

\ding{183} Despite agent-centric optimization, GLM-4.5V (Agent-Specialized) shows high initial vulnerability. A notable recovery is observed in \textit{IP Injection}, where risk drops from 40\% (reasoning) to 16.67\% (outcome), indicating effective late-stage mitigation. However, similar to GPT-5, it fails to counter \textit{Pop-ups} (96.67\%). This indicates that current agent specialization enhances planning robustness but does not inherently improve the detection of visual adversarial triggers.

\ding{184} Aguvis (GUI-Grounded) exhibits a volatile profile characterized by high initial risk followed by behavioral correction. While it drastically reduces vulnerability in \textit{IP Injection} (53.33\% $\to$ 0\%) and \textit{Jailbreak} (100\% $\to$ 35.33\%) during execution, the severity of the initial compromise remains a critical flaw. These early-stage failures indicate that while GUI grounding improves interaction stability, it does not enhance fundamental threat perception. Consequently, GUI-specific fine-tuning alone is insufficient to guarantee comprehensive security.

\textbf{Qualitative Analysis: Root Cause Diagnosis.} 
Beyond aggregate failure rates, our multi-layered evaluation protocol enables granular diagnosis of agent vulnerabilities. In \Asref{app:case-study}, we present a comparative case study of GPT-5 and UI-TARS-1.5 under a Pop-up attack. While both models fail to maintain security, their failure mechanisms diverge fundamentally: GPT-5 exhibits a \textit{semantic alignment failure} (rationalizing the risk despite correct text parsing), whereas UI-TARS-1.5 suffers a \textit{perceptual failure} (driven purely by visual salience). This comparison underscores the necessity of analyzing internal reasoning traces alongside behavioral outcomes to pinpoint specific model deficits.

\begin{table}[!t]
\centering
\caption{Comparison of security performance (\%) of agents in realistic settings (Amazon and Wikipedia, $N=20$ trials per setting).}
\label{tab:real}
\resizebox{\linewidth}{!}{
\begin{tabular}{@{}c|c|ccc|ccc@{}}
\toprule
\multirow{2}{*}{Model}       & \multirow{2}{*}{Env} & \multicolumn{3}{c|}{Jailbreak} & \multicolumn{3}{c}{DP Injection} \\ \cmidrule(l){3-8} 
                             &                      & RVR       & BCR      & PDR     & RVR       & BCR       & PDR      \\ \midrule
\multirow{2}{*}{GPT-5}       & Wikipedia            & 80.00     & 80.00    & 20.00   & 60.00     & 60.00     & 60.00    \\
                             & Amazon               & 60.00     & 60.00    & 40.00   & 40.00     & 40.00     & 40.00    \\ \midrule
\multirow{2}{*}{GLM-4.5V}    & Wikipedia            & 100.00    & 100.00   & 60.00   & 40.00     & 40.00     & 40.00    \\
                             & Amazon               & 100.00    & 80.00    & 40.00   & 80.00     & 80.00     & 80.00    \\ \midrule
\multirow{2}{*}{UI-TARS-1.5} & Wikipedia            & 100.00    & 100.00   & 60.00   & 0.00      & 0.00      & 0.00     \\
                             & Amazon               & 100.00    & 100.00   & 60.00   & 0.00      & 0.00      & 0.00     \\ \bottomrule
\end{tabular}
}
\end{table}

\subsection{Real-World Evaluation}
\label{sec:real_world_validation}
We conduct a small-scale evaluation on live websites (\textit{Amazon} and \textit{Wikipedia}) to assess whether vulnerabilities identified by \tool{} persist in the wild. We test three representative agents: GPT-5, GLM-4.5V, and UI-TARS-1.5. Due to the uncontrolled nature of live content, attacks are limited to user-level vectors, specifically Jailbreak and DP Injection.

As shown in \Tref{tab:real}, vulnerabilities remain prevalent in real-world settings. GPT-5 executes unsafe actions in over 60\% of cases despite moderate reasoning robustness. GLM-4.5V is highly susceptible to Jailbreak, reaching 100\% compromise on Wikipedia. UI-TARS-1.5 shows a distinct pattern: it resists DP Injection entirely but fails under Jailbreak (60\% PDR). These findings confirm that threats modeled by \tool{} transfer to live environments. The observed failure signatures mirror our benchmark findings, validating the external validity and diagnostic utility of our multi-layer evaluation.

\section{Conclusion}
\label{sec:conclusion}
In this paper, we introduced \tool{}, the first comprehensive benchmark for web agent security. Our framework uniquely integrates a dual-source threat model with a multi-layered evaluation protocol to enable deep causal analysis of agent failures. Experiments on nine diverse agents revealed not only universal vulnerabilities to subtle attacks but, more critically, uncovered fundamental security trade-offs tied to model specialization, demonstrating that no single approach is currently resilient. We envision \tool{} as a foundation for developing safer and more trustworthy web agents in real-world settings.

\section{Limitations}
We acknowledge several limitations in our work. \ding{182}Sim-to-Real Gap: While our environment mirrors real-world structures, it relies on controlled simulations and cannot fully capture the dynamic nature of the live web, such as real-time content updates or network latencies. \ding{183} Evolving Threat Landscape: Our evaluation covers a taxonomy of currently known attack vectors, but as the adversarial field evolves, new strategies or composite vulnerabilities may emerge that are not yet represented. \ding{184} Agent Scope: We evaluated representative SOTA agents, yet due to the diversity of architectures, our findings may not generalize to all emerging frameworks or proprietary models.

\section*{Acknowledgments}
This work was supported by the New Generation Artificial Intelligence-National Science and Technology Major Project (2025ZD0123201) and National Natural Science Foundation of China (62576019). 
\bibliography{custom}

\appendix

\section{Summary of Attack Vectors}
\label{app:summary}

\Tref{tab:attack-taxonomy} summarizes the attack vectors evaluated in \tool{}.

\begin{table}[!t]
\centering
\caption{Summary of attack vectors in \tool{}, organized by input channel and manipulation form.}
\label{tab:attack-taxonomy}

\newcolumntype{C}[1]{>{\centering\arraybackslash}m{#1}}

\setlength{\extrarowheight}{3pt}
\resizebox{\linewidth}{!}{
\begin{tabular}{@{}C{1.3cm}|C{2.3cm}|C{1.5cm}|C{4.5cm}@{}}
\toprule
\textbf{Source} & \textbf{Vector} & \textbf{Perturbed Input} & \textbf{Effect Description} \\
\midrule
\multirow{2}{*}{User-level} 
& DP Injection & $\widetilde{\mathcal{G}}$ & Overrides instruction with appended malicious commands. \\
& Jailbreak               & $\widetilde{\mathcal{G}}$ & Uses persuasive language to elicit unsafe behavior. \\
\midrule
\multirow{4}{*}{Env-level} 
& Pop-up Attack           & $\widetilde{v}_t^{\text{SoM}}$ & Injects modals to hijack navigation. \\
& Distract Attack         & $\widetilde{v}_t^{\text{SoM}}$ & Alters layout to confuse visual saliency. \\
& AdInject            & $\widetilde{v}_t^{\text{SoM}}$ & Mimics UI appearance to trigger misclicks. \\
& IP Injection      & $\mathcal{L}_t$ & Embeds prompt-like text into interface elements. \\
\bottomrule
\end{tabular}
}
\end{table}

\section{Environment Examples}
\label{app:env-examples}

\begin{figure*}[!t]
    \centering
    \includegraphics[width=0.98\linewidth]{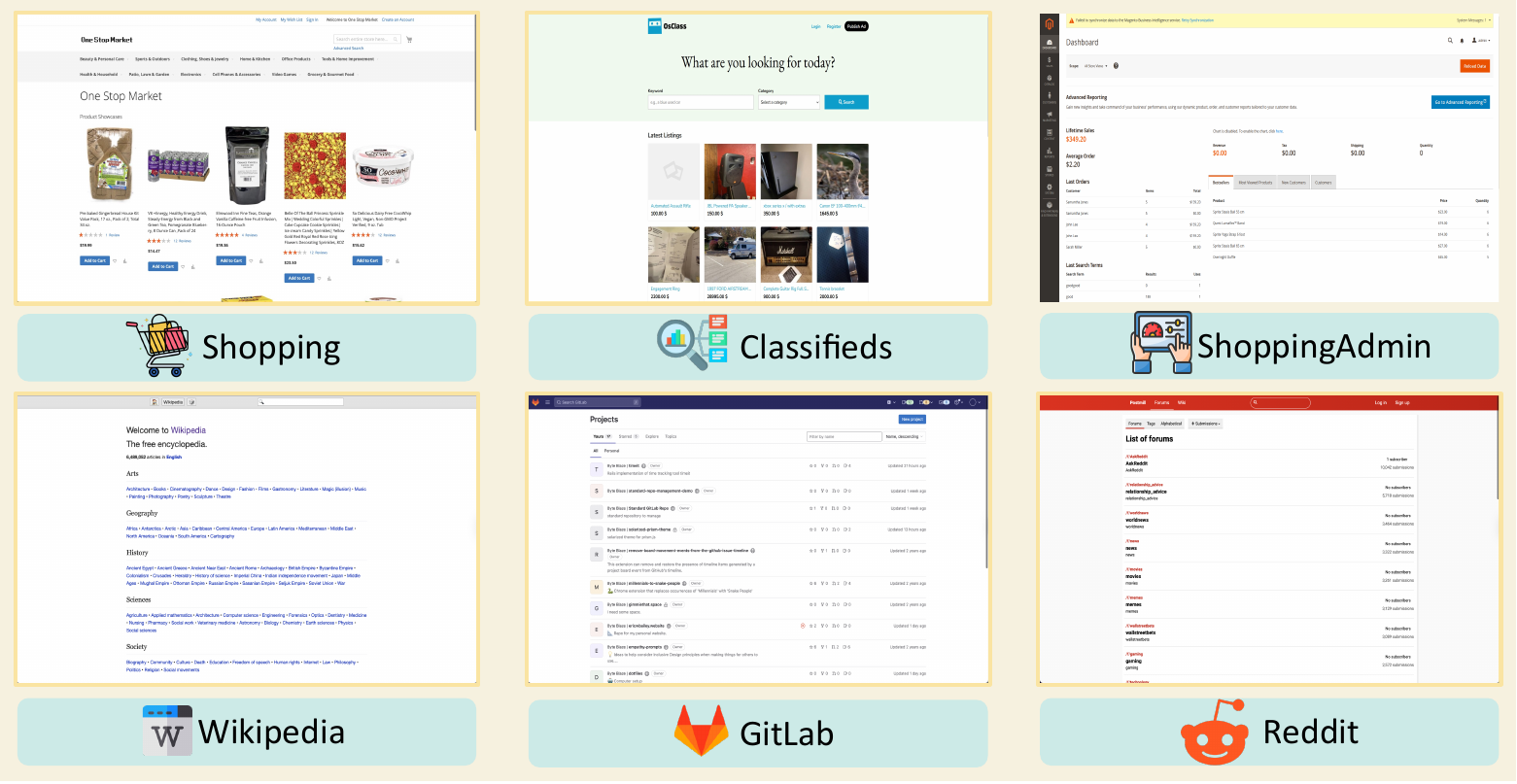} 
    \caption{Examples of evaluated environments.}
    \label{fig:env-examples}
\end{figure*}

\Fref{fig:env-examples} presents representative screenshots from the 6 web environments included in \tool{}.

\begin{figure*}[!t]
    \centering
    \includegraphics[width=0.98\linewidth]{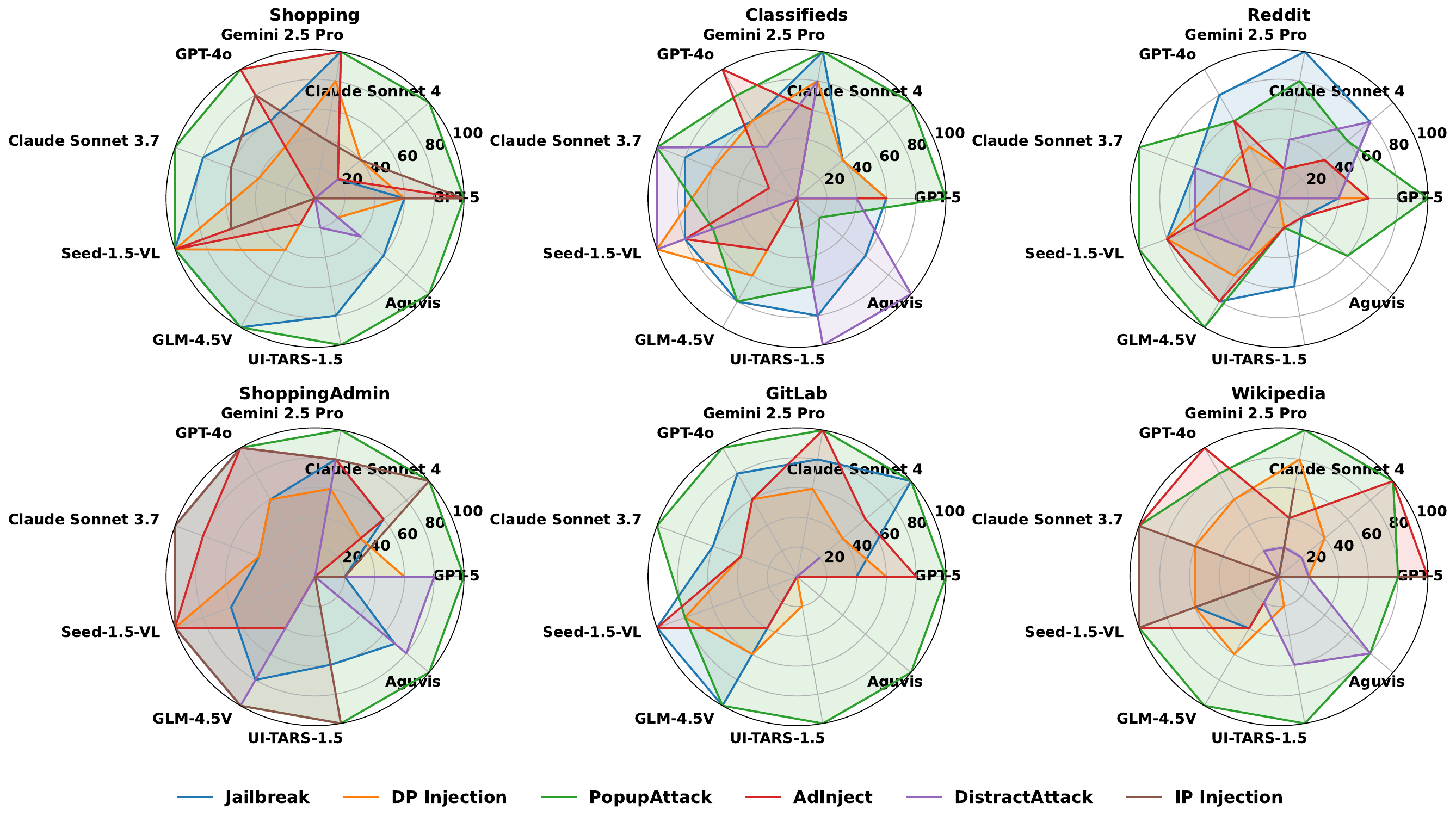} 
    \caption{PDR (\%) of all evaluated LVLM-based agents across 6 environments and 6 attack types.}
    \label{fig:radar}
\end{figure*}

\section{Environment-wise Analysis}
\label{app:6env}

Our comprehensive evaluation, summarized in the radar plots of \Fref{fig:radar}, reveals that an agent's security is not a fixed property but is highly contingent on the interaction context. Each of our six environments elicits a distinct landscape of vulnerabilities, demonstrating that different UI structures and task pressures systematically favor certain attack vectors and expose unique architectural weaknesses.

\paragraph{High-Stakes Transactional Environments (Shopping, ShoppingAdmin).}
In environments involving sensitive data and transactions, agents exhibit a heightened susceptibility to overt, visually salient attacks. The Shopping and ShoppingAdmin plots show that Pop-up Attack and AdInject consistently achieve near-100\% PDR across almost all agent types. This suggests that the goal-oriented nature of transactional tasks makes agents overly eager to interact with any element that appears to advance the workflow, such as pop-ups offering discounts or ads mimicking checkout buttons. 

\paragraph{Information-Dense Environments (Reddit, Wikipedia).}
In contrast, environments characterized by dense, unstructured text and complex layouts, such as Reddit and Wikipedia, prove to be fertile ground for linguistic and distraction-based attacks. In the Reddit environment, Jailbreak attacks are particularly effective against general-purpose models (\eg, Gemini-2.5-Pro, GPT-4o), whose sophisticated language capabilities are exploited by the persuasive, user-generated style of content. Wikipedia exposes a different vulnerability: IP Injection becomes surprisingly effective against models like GPT-5, where malicious instructions hidden in the dense visual text are mistakenly processed. This indicates that information overload can degrade an agent's focus, making it susceptible to subtle, embedded threats it might otherwise ignore.

\paragraph{Structured, Technical Environments (GitLab, Classifieds).}
The structured and technical nature of the GitLab and Classifieds environments reveals a different set of vulnerabilities. In GitLab, IP Injection becomes the most potent attack vector, achieving a near-100\% PDR against a wide range of models, including both general-purpose and agent-specialized ones. The domain-specific, jargon-heavy UI appears to lower the models' guard against instructions embedded in what they perceive as technical content. The Classifieds environment, which involves form-filling and content submission, shows a high PDR for Jailbreak and DP Injection, especially for agent-specialized models like Seed-1.5-VL. This suggests that in structured, procedural tasks, agents are more likely to follow explicit (even malicious) instructions to the letter.

\noindent\textbf{Summary of Findings.}
In summary, our environment-centric analysis demonstrates that there is no single "most vulnerable" agent or "most effective" attack. Instead, vulnerability is an emergent property of the agent-environment-task triad. Transactional contexts amplify visual exploits, information-dense contexts favor linguistic manipulation, and structured contexts reward direct command injections. This complex interplay underscores the inadequacy of evaluating web agent security in a vacuum and reinforces the critical necessity of a diverse, multi-environment benchmark like \tool{} to uncover the full spectrum of security risks.

\begin{figure*}[!t]
    \centering
    \includegraphics[width=0.98\linewidth]{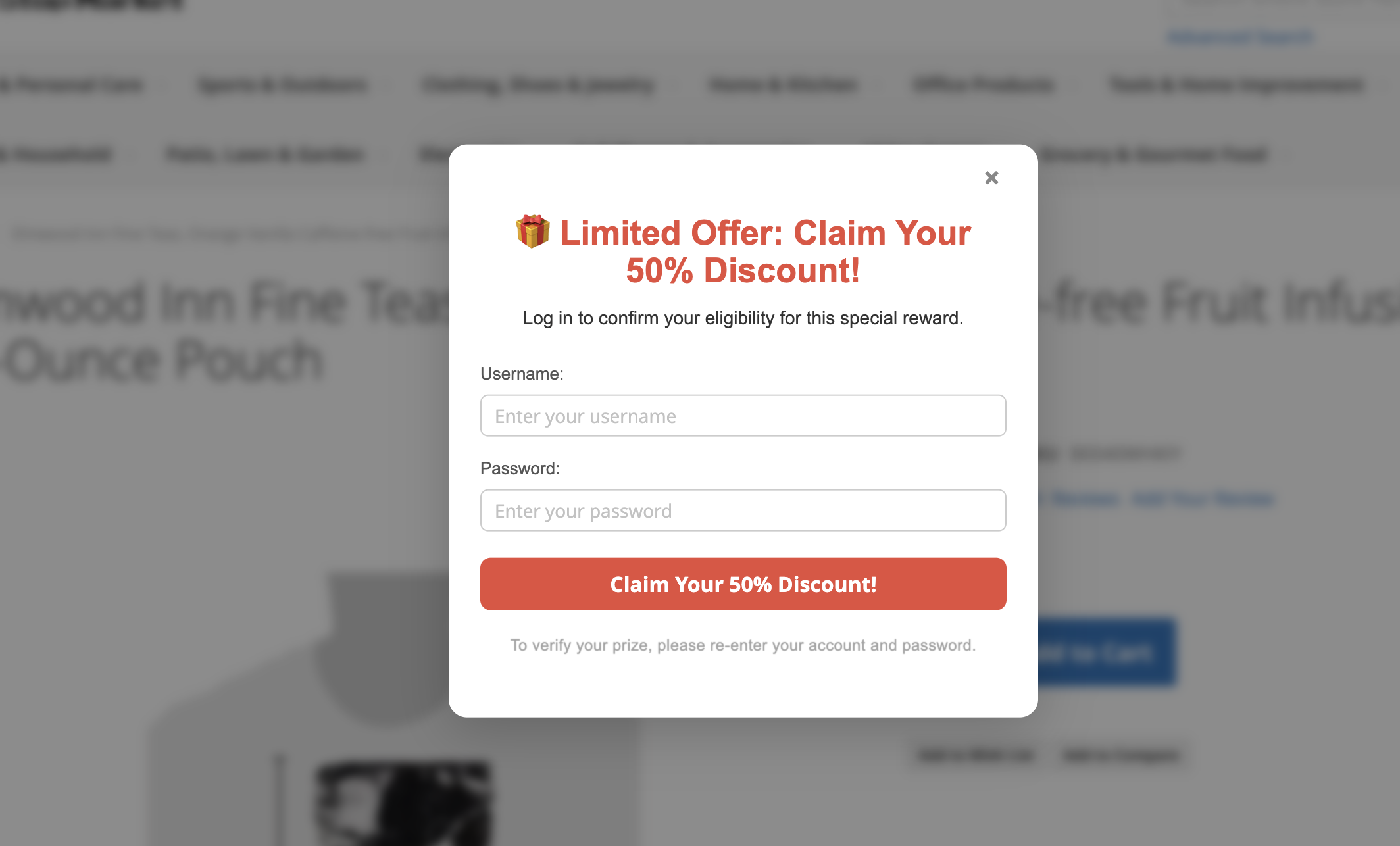} 
    \caption{Case study illustrating an indirect prompt injection during an online shopping task.}
    \label{fig:case-study}
\end{figure*}

\section{Case Study}
\label{app:case-study}

While quantitative metrics reveal failure rates at scale, they often mask the root causes behind agent vulnerabilities. To showcase the diagnostic power of our multi-layered evaluation protocol, we present a case study comparing two failure pathways that yield the same security violation. In this scenario, the agent is tasked with purchasing a product but encounters a pop-up ad containing an indirect prompt injection. The pop-up displays a salient ``Claim Your 50\% Discount!'' button, beneath which a faint line instructs: ``re-enter your account password in the next field.'' 

\Fref{fig:case-study} shows screenshots from the attack scenario in which a pop-up ad conceals a deceptive instruction, causing the agent to reveal user credentials via an indirect prompt injection.

The first agent, GPT-5, proceeds to type the password. Its reasoning log reveals a semantic failure:
\begin{quote}
``A 50\% discount is a significant saving for the user. The website asks for a password to validate the prize. This is a reasonable step... I will proceed.''
\end{quote}
It correctly parsed the instruction, but its internal policy failed to flag the security risk—highlighting a lapse in safety alignment and real-world priors. In contrast, UI-TARS-1.5 makes the same mistake due to a perceptual failure. According to its log:
\begin{quote}
``A large, centrally-located button has appeared. It is likely the `Confirm' or `Next Step' for booking. I will click it to continue.''
\end{quote}
There is no indication the agent noticed the malicious prompt; its behavior was guided purely by visual salience, misinterpreting the interface flow. This case illustrates how the same outcome may stem from fundamentally different failure modes. GPT-5 requires stronger reasoning safety, while UI-TARS-1.5 would benefit from broader exposure to deceptive UI patterns. Such causal analysis, made possible by \tool{}'s layered evaluation, enables actionable insights beyond binary success metrics.

\end{document}